\documentstyle[12pt,epsf,amsfonts,amssymb]{article} 
\input amssym.def
\input amssym.tex

\DeclareFontFamily{U}{rsfs}{\skewchar\font"7F}
\DeclareFontShape{U}{rsfs}{m}{n}{
	<-6> rsfs5
	<6-8> rsfs7
	<8-> rsfs10
	}{}
\DeclareMathAlphabet{\mathscr}{U}{rsfs}{m}{n}

\def\ben{\begin{equation}}
\def\een{\end{equation}}
\def\bea{\begin{eqnarray}}
\def\eea{\end{eqnarray}}

\newcommand{\I}{{\mathscr I}} 
  
\newcommand{\non}{\nonumber}    

\begin{document}

\title{Topology and Signature Changes in Braneworlds} 
\author{Gary W. Gibbons and Akihiro Ishibashi \\
{\it D.A.M.T.P., University of Cambridge, } \\ 
{\it Wilberforce Road, Cambridge CB3 0WA, U.K. } } 

\maketitle

\begin{abstract} 

It has been believed that topology and signature change of the universe 
can only happen accompanied by singularities, in classical, 
or instantons, in quantum, gravity. 
In this note, we point out however that 
in the braneworld context, such an event can be understood 
as a classical, smooth event. We supply some explicit examples of 
such cases, starting from the Dirac-Born-Infeld action. 
Topology change of the brane universe can be realised by allowing 
self-intersecting branes. Signature change in a braneworld 
is made possible in an everywhere Lorentzian bulk spacetime. 
In our examples, the boundary of the signature change is a 
curvature singularity from the brane point of view, but  
nevertheless that event can be described in a completely 
smooth manner from the bulk point of view. 

\end{abstract}    

\section{Introduction}

Since the demonstration that gravity can be localised on 
a $(p+1)$-dimensional submanifold of a higher dimensional 
bulk spacetime~\cite{RS2}, there has been a revival of 
interest in the old 
idea~\cite{old1,old2,old3,old4,old5,old6,old7,old7-1,old7-2,old7-3,old8} 
that rather than taking the 
higher dimensional spacetime to be a (possibly warped) product of
a compact \lq internal space' with a four-dimensional spacetime 
one should instead regard our spacetime as a 3-brane embedded, or
more speculatively immersed (i.e having self-intersections),
in a bulk spacetime. Once this idea has been accepted it becomes
natural to ask whether such branes can collide, as in the ekpyrotic
scenario~\cite{KOST,KKL}, or oscillate cyclically~\cite{cyclic}. 
It is clear that models of this kind entirely change the context 
in which old problems in cosmology, such as the singularity theorems, 
the possibility of topology change, 
the birth of the universe from nothing etc should be discussed.
In particular the idea that it is sufficient to consider
our universe as a purely self-contained four-dimensional
Lorentzian  spacetime whose evolution can be discussed without 
reference to the bulk becomes untenable. Moreover in violent
processes, such as brane collisions, one expects the usual clear-cut 
distinction between brane and bulk to break down. 
We have also known for many years~\cite{Geroch,Tipler} 
that if changes of topology are involved, spacetime cannot be 
causal and time orientable and admit an everywhere smooth non-singular 
Lorentzian metric. So far, the main response to this obstacle 
has  either been to adopt singular 4-dimensional  Lorentzian metrics,  
or to appeal to quantum processes mediated by gravitational
instantons\footnote{ 
Instantons describing a vacuum bubble nucleation on a Randall-Sundrum 
type braneworld have been discussed in \cite{GP2002}. 
}, 
that have Riemannian\footnote{ 
In this paper we shall use the word {\it Riemannian} for any positive 
definite metric. Often in the physics literature, especially in
connection with the so-called Euclidean approach, such metrics 
are referred to as {\it Euclidean}, even though they may not be flat. 
To avoid confusion, we prefer to adhere to the standard mathematical 
nomenclature. We also use ${\Bbb E}^{p,q}$ to denote ${\Bbb R}^n$, $n=p+q$, 
equipped with a flat metric of signature $(p,q)$. 
Thus ${\Bbb E}^n$ is $n$-dimensional Euclidean space. 
} 
metrics. One view point on that latter approach is to consider spacetimes 
admitting a change of signature, the instanton being regarded as a region
of spacetime where the spacetime signature is $++++$. 
It seems reasonable to expect that a successful theory
of brane collisions should throw much needed light on these, 
at present rather obscure, issues.   

\par 
Modelling the collision of branes using the full equations of motion 
of the bulk spacetime\footnote{
Besides the ekpyrotic scenario~\cite{KOST,KKL}, 
there have been works on brane collisions in the Randall-Sundrum 
type models (see e.g.,~\cite{Perkins,Bucher,GIT,KSS}).   
}, even supposing a classical or semi-classical approximation 
to be valid, is technically an extremely challenging 
task and so far comparatively small progress has been made, and much
of it in the adiabatic approximation in which the collision
process is supposed to be slow. One then considers a rather 
conventional four-dimensional cosmology of the Friedman-Lemaitre 
type  with additional scalar fields 
representing the separation of the two colliding branes.
The associated scalar fields are very similar to, and in some
cases may be identified with, the tachyon fields encountered in open
string theory. In these approaches gravity is fully taken into
account, but not the loss of distinction between brane and bulk, 
nor are issues of singularities, topology and signature change 
fully faced up to. 
Signature changes in the Randall-Sundrum braneworld context 
have been discussed in \cite{MSV}. 

In this note we wish to explore a different, and hopefully
complementary approximation which,
while gravity is ignored, provides a perfectly smooth and
non-singular account of both topology and signature change. 
A further merit of our description is that it is extremely simple. 
The basic starting point is the Dirac-Born-Infeld 
action for a D-brane, which has previously 
proved so effective  in tackling global questions
of this kind in String/M-theory. In what follows we shall set to zero
the Born-Infeld world volume gauge field and tachyon field 
and consider only as an action functional 
the total area or volume of the brane 
\ben
 - T_p A(\Sigma _{p+1} ) 
 = -T_p \int_{\Sigma_{p+1}}
   \sqrt{ \left| \det (g_{\mu \nu} \partial_a X^\mu 
          \partial_b  X^\nu) \right|} 
   d^{p+1} u^1 u^2 \dots u^{p+1},  
\label {action} 
\een 
where $u^a$, $a=1,2,\dots, p+1$ are world volume coordinates of 
the $p$-brane $\Sigma _{p+1}$, and $X^\mu = X^\mu (u^a)$, 
$ \mu = 0, 1, \dots, n$ gives an immersion of $\Sigma_{p+1} $ 
into an $(n+1)$-dimensional spacetime with metric $g_{\mu \nu}$.  
The constant $T_p$ is the brane-tension and its value will play 
no role in what follows. 
Some very interesting solutions of the equation of motion including 
Born-Infeld field were obtained in Refs.~\cite{HHW,HHNW}. 
These solutions exhibit signature change. The result we are about 
to describe shows that the Born-Infeld field is not necessary for 
signature change. In the case when the tachyon and Born-Infeld field 
are non-vanishing, there is a question of which metric, the open
string metric or induced metric~\cite{GH2001} 
undergoes signature change. 
Some ideas about signature change at finite temperature in external 
electromagnetic field for open strings may be found
in~\cite{KarPanda}. 
In the present case, there is only one metric and this question 
does not arise for us. 

\par 
The Euler-Lagrange equations are easily  
derived,  
\ben
  \partial_a \left(\sqrt{\det(h_{cd})}h^{ab}\partial_b X^\mu \right) 
   - \sqrt{\det(h_{cd})}\Gamma^{\mu}_{\nu \lambda} 
    h^{ab} \partial_a X^\nu\partial_b X^\lambda 
  = 0 ,         
\label{E-L:eom}
\een
where $h^{ab}$ is the inverse of 
$h_{ab} = g_{\mu \nu} \partial_a X^\mu \partial_b  X^\nu$ and 
$\Gamma^{\mu}_{\nu \lambda}$ is the Christoffel symbol with 
respect to $g_{\mu \nu}$. 
Geometrically the variational equation amount to the statement 
that the mean curvature vector, 
i.e., the vector obtained by taking the trace of the $(n-p)$  
second fundamental forms associated with the co-dimension 
$n-p$ immersion, vanishes. In what follows we shall restrict ourselves 
to the case of a hypersurface for which $n-p=1$ and the physical case 
is thus $p=3,\, n=4$. 
If spacelike, and immersed in a Riemannian space, 
hypersurfaces of this type are often called \lq minimal', 
even though in general they will only be saddle points of the area 
or volume functional. It makes no sense to speak of 
`maximal' submanifolds of a Riemannian manifold because 
almost all variations will be area increasing. 
Hypersurfaces which are true local minima are often 
referred to as `stable'. By contrast a spacelike hypersurface 
in a Lorentzian manifold 
is said to be maximal. This is because almost all 
variations of a spacelike hypersurface of a Lorentzian
spacetime will decrease its volume.  Maximal hypersurfaces
have been extensively studied in the mathematical and general
relativity literature. The case we are interested in is the timelike
case  which have, by comparison, hitherto been rather neglected. 
Such timelike hypersuraces may also be called maximal. 

The action (\ref{action}) has a great deal of gauge-invariance
(${\rm diff}(\Sigma _{p+1})$). We fix it by adopting what is 
sometimes inaccurately called `static' gauge. 
A better term is Monge gauge~\cite{Monge}. It amounts to using 
a height function as the basic variable. One could choose
the height coordinate to be spacelike or timelike, independently 
of whether the hypersurface is timelike or spacelike. We shall choose
a {\sl timelike height function}  $X^0=t(u^a)$ to specify our
hypersurface, with $X^\alpha =u^a$, $\alpha = 1, \dots, n=p+1$. 
For most cases in this note, $g_{\mu \nu}$ is taken as the Minkowski 
metric $\eta_{\mu \nu}$ and only the first term 
of eq.~(\ref{E-L:eom}) is relevant. 
   
{\sl If the hypersurface is spacelike or timelike, the equation of motion
is the same}, it is 
\ben
 \left\{ 1- \delta^{cd} (\partial_c t) \partial_d t \right\} 
 \delta^{ab}\partial_a\partial_b t 
  + \delta^{ac}\delta^{bd}(\partial_a t)( \partial_b t )
    \partial_c \partial_d t =0. 
\label{eom} 
\een

It is perhaps worth re-emphasising that eq.~(\ref{eom}) is valid 
for both spacelike or timelike hypersurfaces and it remains well
defined if the signature of the metric 
$h_{ab}=\delta_{ab}- (\partial_a t)\partial_bt$ 
induced on the hypersurface changes sign, that is 
if $\delta^{ab}(\partial_a t)\partial_b t$ passes 
through unity. This strongly indicates that eq.~(\ref{eom}) 
presents no obstacle to signature change,  
a fact we shall confirm in detail shortly. Although eq.~(\ref{eom})
has been derived from the Dirac-Born-Infeld 
action (\ref{action}), we have eliminated
square roots, which is essential to allow a smooth signature change.
To emphasise this point, we shall refer to eq.~(\ref{eom}) as the
(Lorentzian) Laplace-Young equation since it is the Lorentzian
analogue of the Laplace-Young equation which arises in the study of
soap films and other areas of condensed matter physics---see e.g., 
\cite{KL99,K00}. 
It may be regarded as a non-linear equation for $t(u^a)$ 
with $p+1$ independent variables which may change from hyperbolic 
to elliptic,  
when the matrix 
\ben
\delta_{ab} \{ 1- \delta^{cd}(\partial_c t) \partial_d t\} 
         + (\partial_a t) \partial_b t 
\een       
changes signature. This occurs precisely when
\ben
\delta^{cd} (\partial_c t) \partial_d t = 1.  
\een 
Thus strictly speaking, even in the Lorentzian part of the world
volume, one cannot speak of a Cauchy problem for the evolution. 
It is this feature of the equation which permits signature 
and topology change. 

One strategy for obtaining solutions of eq.~(\ref{eom}) is to take a 
known explicit solution $z=z(u^a)$ of the standard minimal 
surface equation for a spacelike minimal surface in flat 
Euclidean space and \lq Wick rotate', that is analytically continue 
it in such a way that $z \rightarrow it$.

\section{Self-Intersecting Branes and Topology Change} 

\subsection{Lorentzian Enneper Solution} 

For our first example we start with Enneper's surface~\cite{Enneper} 
which is given by 
\ben  
x= 3 \alpha + 3 \alpha \beta^2 -\alpha^3,
\een
\ben
y= 3 \beta +  3 \alpha^2 \beta -\beta^3,
\een
\ben
z=3 \alpha ^2 - 3 \beta^2, 
\een
where $\alpha$ and $\beta$ are world volume coordinates 
in conformal gauge. 
The metric induced on the surface is
\ben
ds ^2 =9 ( 1 + \alpha^2 + \beta^2 )^2 ( d \alpha^2 + d \beta^2 
).\een

The Wick rotation consists of setting $\alpha =\sigma$, $\beta =i
\tau$
with $\sigma $ a real spacelike coordinate and
 $\tau$ a real timelike coordinate.
This results in   $y=it$ with $t$ real and relabelling
$z$ as $y$ we get
\ben
x=  \sigma (3 - 3 \tau^2  -\sigma^2) , 
\een 
\ben
y=3 \sigma ^2 + 3 \tau^2 , 
\een
\ben
t= \tau( 3 +   3 \sigma^2 + \tau^2). 
\een

The induced metric is
\ben
 ds^2 = 9(1 + \sigma^2 -\tau^2 )^2 ( d \sigma ^2 -d\tau^2). 
\een 
A trivial extension to the $4$-dimensional case is obtained by adding 
$dz^2+dw^2$ to the above metric. The surface is time symmetric 
in the sense that $\tau \rightarrow -\tau$ 
takes $t \rightarrow -t$. 
If $ \sigma$ and $\tau$ run over ${\Bbb R}^2$ we have an everywhere 
Lorentzian metric except on the null curves 
$\tau=\pm \sqrt{1+ \sigma^2}$ 
at which the conformal factor vanishes. A simple calculation reveals 
that the $3 \times 2 $ Jacobian matrix 
$\partial (x,y,t) / \partial (\sigma, \tau)$ 
has rank 2 except on the two null curves. This implies that we have 
a smooth immersion or embedding away from the two null curves. 
The formula for $x$ as a function of $\sigma$ 
at fixed $\tau$, is a cubic, anti-symmetric in $\sigma$ which always 
passes through $x=0$ at $\sigma =0$. For $ 0\le \tau ^2 <1$ 
the cubic has two other roots at $\sigma = \pm \sqrt 3 \sqrt {1-\tau ^2 }$.
Thus in this interval, $\sigma$ is not a single valued function 
of $x$. For $\tau = \pm 1$, the three roots coalesce at the beginning
of the null curves $\sigma = \pm  \sqrt{\tau ^2 -1 }$, $\tau^2>1$, 
and there after $\sigma$ is a single valued function of $x$. 
In fact eliminating $\sigma$ we find
\ben
x= \pm \sqrt{ \frac{y}{3}- \tau ^2 }\left
       (3-2\tau^2-\frac{y}{3} \right). 
\een

\begin{figure}[h] 
 \centerline{\epsfxsize = 8.0cm \epsfbox{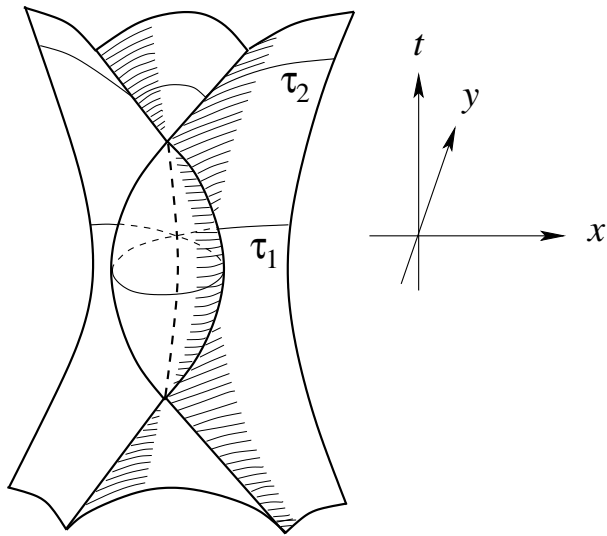}}
\vspace{3mm}
\begin{center}
\begin{minipage}{14cm}
     \caption{\small The world volume geometry of Enneper's surface. 
             The $(z,w)$ dimensions are suppressed. The dashed thick line
             denotes the self-intersection points on the brane. 
             Thin lines (surfaces) describe $\tau=const.$ surfaces. 
             The line has a loop at $\tau_1 \,(0 \le \tau_1 <1)$, 
             and, at $\tau_2 \, (>1)$, it forms two kinks moving 
             outward along null curves.              
             } 
        \protect \label{fig:Ennerper}
\end{minipage}
\end{center}
\end{figure}

The interpretation is that, neglecting the coordinates $z, w$
 we have an infinitely
long piece of string, symmetric about the $y$ axis  which intersects 
itself on the $y$ axis at $y=\sqrt 3 \tau $ for $0\le \tau ^2 <1$. 
This loop tightens up and shortens forming a kink which then 
moves outward along the null curve at the speed of light as 
illustrated in Figure~\ref{fig:Ennerper}. 
The world sheet is a smooth immersion, rather than an embedding 
except precisely on the kink. This describes a self-intersecting 
braneworld that changes its spatial topology from a loop to a cusp 
and then a kink.   

\subsection{Self-intersecting solutions of $S^{m-1} \times dS_{n-1} $-type} 

{}Further examples of smoothly  intersecting branes may be obtained 
by adapting and extending the work of~\cite{Allencar}. 
He constructs Riemannian minimal $(2m-1)$-dimensional 
submanifolds of  ${\Bbb E} ^m \times {\Bbb E}^m$ invariant under 
$SO(m) \times SO(m)$ and shows that generically they intersect
on $S^{m-1} \times S^{m-1}$. Geometrically the submanifolds
are warped product of the form ${\Bbb R} \times S^{m-1} \times
S^{m-1}$.
  
These would of course provide static intersecting $p$-branes with 
$p=2m-1$. One may generalise his set up slightly by considering 
$(2m-1)$-dimensional minimal submanifolds of 
${\Bbb E} ^m \times {\Bbb E} ^{n-1,1}$ invariant under 
$SO(m) \times SO(n-1,1)$. One sets 
\ben
X^\mu = ( x(t) {\bf n}_m ,y(t) {\bf n}_n),  
\een
where ${\bf n}_m$ is a unit vector in ${\Bbb E} ^m$ defining 
a unit $S^{m-1}$ and ${\bf n}_n$ is a unit spacelike vector 
in ${\Bbb E} ^{n-1,1}$ defining a de Sitter space $dS_{n-1}$ of unit
radius.  The equations of motion for $x(t)$ and $y(t)$ may be obtained
by substituting in the Dirac-Born-Infeld 
action (\ref{action}) to get a Lagrangian 
\ben
L= x^{m-1} y ^{n-1} \sqrt{{\dot x } ^2 + {\dot y}^2 }. 
\label{lagrange} 
\een
Some calculations lead to the equation 
\ben
{\dot y} {\ddot x} - {\dot x} {\ddot y} 
= { m+n-2 \over 2} \Bigl( { {\dot y}
  \over x}  - { {\dot x} \over y} \Bigr) ({\dot x} ^2 + {\dot y} ^2 ).
\een

If $m=n$ this coincides with the equation obtained
in~\cite{Allencar}. At this stage we still have the freedom to
choose the parameter $t$. One choice (adopted in~\cite{Allencar}) 
is to use arc length, i.e. set
\ben
{\dot x} ^2 + {\dot y} ^2 =1. 
\label{condition}
\een 
If $m=n$,  the analysis  proceeds exactly as in~\cite{Allencar}.
A glance at figures (a) and (b) in~\cite{Allencar} reveals 
the existence of multiply intersecting solutions 
(for which $x=y$ and for which $x$ and $y$ never vanish)
 of topology ${\Bbb R} \times S^{m-1} \times dS_{m-1}$ and
multiply intersecting  solutions for which $x$ vanishes at one end
point, of  topology  ${\Bbb R} ^m \times dS_{m-1}$.
   
Despite the self-intersections, the intrinsic geometry on the
$p$-brane is everywhere smooth
and the metric everywhere Lorentzian. This is related to the fact that
we adopted the gauge  condition (\ref{condition}) which forbids
signature change. 
 
It is interesting to speculate about what happens near
the self-intersection surface. As far as the world volume 
theory is concerned this is a smooth timelike hypersurface, 
a sort of domain wall at which four spacetime regions meet. 
One question is whether one 
can pass  from one sheet to another. One might also
wonder whether charges which would otherwise be conserved
can leak from one sheet to another.

\section{Cyclic and Spinning Braneworlds} 

\subsection{Lorentzian Catenoid and Helicoid} 

We first recapitulate the two-dimensional world sheet case. 
Perhaps the most familiar minimal surface is the catenoid.
This takes the form
\ben
\sqrt {x^2 + y^2}= \cosh z. 
\een
Wick-rotating gives the world sheet of a collapsing 
circular loop of string 
\ben
  \sqrt{x^2 + y^2 } = \cos t. 
\een 
This corresponds to  the embedding
\ben 
x=\cos \tau \cos \sigma , 
\een
\ben
y=\cos \tau \sin \sigma , 
\een
\ben
t=\tau , 
\een
with induced metric
\ben
ds^2 = \cos ^2 \tau (\ d \sigma ^2 -d \tau ^2 ).  
\label{metric:cyclic}
\een
The embedding is singular at $ \tau = \pm \pi/ 2 + 2 \pi n$, 
$ n \in {\Bbb Z}$. 

It is well known that the catenoid is locally  isometric to the helicoid. 
This corresponds to the spinning string solution for which 
\ben
x= \sin \sigma \cos \tau , 
\een
\ben
y= \sin \sigma \sin \tau , 
\een
\ben
t= \tau , 
\een
with induced metric
\ben
ds^2= \cos^2 \sigma (d \sigma ^2 - d \tau ^2 ).
\een
The ends of the 
string are at $\sigma ={\pi \over 2}$ which move
on a lightlike helix.

The spinning solution and the collapsing circular loop
are related by the interchange of $\sigma $ and $\tau$.
A similar discrete symmetry links the catenoid and helicoids.
In that case, the discrete symmetry is contained in a more general 
$SO(2)$ symmetry discovered in the case 
of usual minimal surfaces by Bonnet~\cite{Enneper} 
which allows one to construct a one parameter family of minimal,
locally isometric, embeddings
connecting the catenoid and the helicoid.

The case of minimal Lorentzian surfaces is slightly different.
There is now a continuous $SO(1,1)$ symmetry, but it does not contain 
the discrete interchange of $\sigma$ and $\tau$. Explicitly 
the following one-parameter family of embeddings is both 
minimal and isometric
\ben
x=\cosh \lambda \cos \sigma \cos \tau 
  - \sinh \lambda \sin \sigma \sin \tau , 
\een 
\ben
y= \cosh \lambda \sin \sigma \cos \tau 
  + \sinh \lambda \cos \sigma \sin \tau , 
\een  
\ben
 t= \sigma \sinh \lambda + \tau \cosh \lambda ,  
\een
where $\lambda$ denotes the $SO(1,1)$ parameter. The induced 
metric becomes the same as~(\ref{metric:cyclic}), and the 
embedding is singular at $\tau = \pm \pi/2 +2\pi n$, $ n \in {\Bbb Z}$ 
irrespective of the value of $\lambda$.  
Clearly $\lambda =0$ is the collapsing circular loop and 
$\lambda \rightarrow \infty$ gives a one-dimensional null helix 
\ben
x=\cos (\sigma + \tau) , 
\een  
\ben
y= \sin(\sigma + \tau) , 
\een
\ben
t=\sigma + \tau. 
\een

{}From the point of view of string theory, this symmetry is 
related to T-duality.  
This works as follows. Consider first the case of a world sheet 
which has positive definite signature \cite{osserman}. 
We introduce isothermal coordinates $z=x^1+ix^2$, 
in which the induced metric is conformally flat
\ben
ds^2 =\Omega ^2 dz d \bar z.
\een
The equation of motion is
\ben
\nabla ^2 X^a=0 \,\, \Rightarrow  \,\,  
{\partial^2  \over \partial z \partial \bar z} X^a=0. 
\een
Thus we may take  $X^a(z,\bar{z})$ to be the real part of some 
holomorphic function. If 
\ben
\phi^a  ={\partial X^a \over \partial z} 
        = \frac{1}{2}
          \left(
                {\partial X^a \over \partial  x^1} 
                - i {\partial X^a \over \partial x^2}
          \right) , 
\een
then the conformal gauge condition is
\bea
  \phi^a \phi^a=0 . 
\label{nullcond} 
\eea
Thus Weirstrass's procedure works as follows:
we obtain a  holomorphic solution of (\ref{nullcond}) and set 
\ben
X^a = \Re \int \phi^a(z) dz .
\een

Now the Bonnet rotation 
\ben
\phi^a(z) \rightarrow e^{i\lambda } \phi^a(z) 
\een
with the angle $\lambda $ constant, leaves (\ref{nullcond}) unchanged 
and preserves holomorphicity. However, it changes the embedding 
but not the induced metric, since $\phi^a \bar \phi^a $ is unchanged. 
Now $z=\sigma +i\tau$, and a Bonnet rotation rotates 
$\partial_\tau X^a$ into $\partial_\sigma X^a$. 
This is essentially T-duality as understood by string theorists 
since it typically rotates Dirichlet into Neumann boundary
conditions. 
  
{}For Lorentzian world sheets a similar procedure will work 
but the Bonnet rotation becomes a Bonnet boost. 
Formally, at least, the easiest way to see this is to replace 
the complex numbers by the `double' or `para-complex' 
numbers. That is one sets 
\ben
  z= x^1+e x^2,\qquad \bar z= x^1-e x^2, \qquad 
{\rm with} \,\,\, e^2 = 1.
\een

\subsection{Generalised Spinning Braneworlds} 
Spinning $p$-brane solutions for $p>2$ have been constructed
which spin in $p$ orthogonal 
$2$-planes~\cite{Kikkawa-Yamasaki,Hoppe-Nicolai}. 
One has $X^0=t$ and 
\ben
X^a = ( f_1(u^1,\dots u^p) \exp {i \omega_1 t}, \dots 
        f_p(u^1, \dots u^p)  \exp{i \omega _p t}, 0,\dots, 0),  
\een
where $f_a(u^b)$ are arbitrary functions and on the boundary of the
$p$-brane
one has 
\ben
\sum_a \omega_a^2 f^2_a |_{\partial \Sigma_{p+1} } =1.
\een

It is  a striking fact that as in the case of a spinning string, 
the induced metric is {\sl static}:
\ben
ds^2 = -dt^2 \Bigl( 1-  \sum_a \omega^2_a f^2_a \Bigr)
      + \sum_a (d f_a) ^2 .
\een 
In other words, despite the fact that the $p$-brane is rotating,
an observer of the world volume would not be aware of any
velocity dependent Corioli effects.  The main  signature of rotation
on the world volume would be a non-trivial Newtonian potential 
giving rise to \lq fictitious' centrifugal forces coming from 
the metric component $h_{tt}$.

\subsection{Generalised Catenoid} 

The catenoid may easily be generalised to $(p+2)$-dimensions. 
The Wick rotation gives an $SO(p+1)$ invariant $p$-brane
whose induced metric is of $k=+1$ 
Friedmann-Lemaitre form. This has been discussed 
previously~\cite{Neilsen,Gibbons}.
As in the case of $p=1$ we obtain a cyclic spacetime
which passes through a spacetime singularity of Big Bang or Big
Crunch type. Using the results of~\cite{Hoppe-Nicolai} 
one may avoid the singularity. 
One has an embedding into ${\Bbb E} ^{2d+2,1}$  
\ben
X^\mu = \bigl (t, r(t) \cos \phi(t)  {\bf n } (u^a), r(t)  \sin \phi
(t)
 {\bf n}(u^a)   \bigr ) \label{Hermann},
\een
where ${\bf n} (u^a)$ is a minimal $p$-surface in a unit $S^d \subset
{\Bbb E} ^{d+1}$. Thus ${\bf n}(u^a) $ is a unit $d+1$ vector 
which satisfies
\ben
 - \nabla^2 _g {\bf n}=p {\bf n} , 
\label{minimal} 
\een 
where $\nabla^2_g$ is the Laplacian on the minimal 
$p$-surface. 
Thus for example, if $p=3$ and $d=4$  one could choose an equatorial 
$S^3$  in $S^4$ to get a 3-brane moving in 11  spacetime dimensions.
However there are other possible  minimal 3-manifolds in $S^4$ available.

The time dependence of $r$ and $\phi$ follows from the Lagrangian
\ben
L= {1\over 2} \Bigl({\dot r} ^2 +r^2 {\dot \phi } ^2 \Bigr) 
   -{ 1\over 2} \Bigl( {r\over c} \Bigr) ^{2p},
\een
where $c$ is an integration constant subject to the constraint that
the energy  takes a particular value 
\ben
{1\over 2} \Bigl({\dot r} ^2 + {h^2 \over r^2} \Bigr) 
 + {1 \over 2} \Bigl({r \over c}\Bigr)^{2p} = {1 \over 2},    
\een
where we have made use of the conservation of $\phi$ momentum 
\ben
h=r^2 { \dot \phi }.
\een
It is helpful to define a (strictly positive) effective potential 
for this one dimensional motion.  
\ben
V_{\rm eff}(r)= {1 \over 2} { h^2  \over r^2} + {1 \over 2} 
 \Bigl({r \over  c}\Bigr)^{2p}.  
\een
It is clear that as long as $h\ne 0$, 
 $r$ oscillates between a minimum and maximum value 
while $\phi$ increases monotonically. In particular $r$ never goes to zero.

The induced metric is
\ben
 ds^2 = -\left( 1- {\dot r} ^2 - r^2 {\dot \phi} ^2 \right)dt^2 
        + r^2(t) d {\bf n}^2. 
\label{general:catenoid}
\een
The quantity $d{\bf n}^2 $ is the metric induced on the 
$p$-brane from its minimal embedding into $S^d$. 
It is the same metric as that used to construct 
the Laplacian in (\ref{minimal}). 

Using the energy constraint we deduce that the induced metric 
is of generalised Friedman-Lemaitre-Robertson-Walker form.
\ben
  ds^2 = - 2 V_{\rm eff} (r(t) ) dt^2 + r^2(t) d{\bf n}^2 . 
\label{FLRW}
\een
This becomes clearer if one introduce a propertime variable $\tau$ by 
\ben
d \tau = dt \sqrt { 2 V_{\rm eff} (r(t) )},
\een 
and casts (\ref{FLRW}) in the form 
\ben
ds^2 = -d \tau ^2 + r^2(\tau) d {\bf n} ^2. 
\een

The examples given above describe an oscillating universe, 
but are, for the purposes of the present paper, not so very 
interesting since neither topology nor signature change takes place.

\section{Braneworlds from Euclidean to Lorentzian} 
\subsection{Generalised Scherk Braneworlds}  
The supply of explicit exact solutions to eq.~(\ref{eom}) 
is not so large, even after 200 years of effort, 
but one stands out, due to Scherk, it is of the form 
$t=f(x) +g(y)$.   
A particular solution is 
\ben
  t= \log( \cosh x) -\log (\cosh y), 
\label{soln} 
\een
where we relabel $X^1=u^1=x$ and $X^2=u^2=y$, $X^3=u^3=z$ and $X^4=u^4=w$.
Note that the solution is defined 
$ \forall x,y,z,w \in {\Bbb R}^4$, and that this surface is invariant 
under the involution 
\ben
(t,x,y) \rightarrow (-t, y, x). 
\een 

\begin{figure}[h] 
 \centerline{\epsfxsize = 12cm \epsfbox{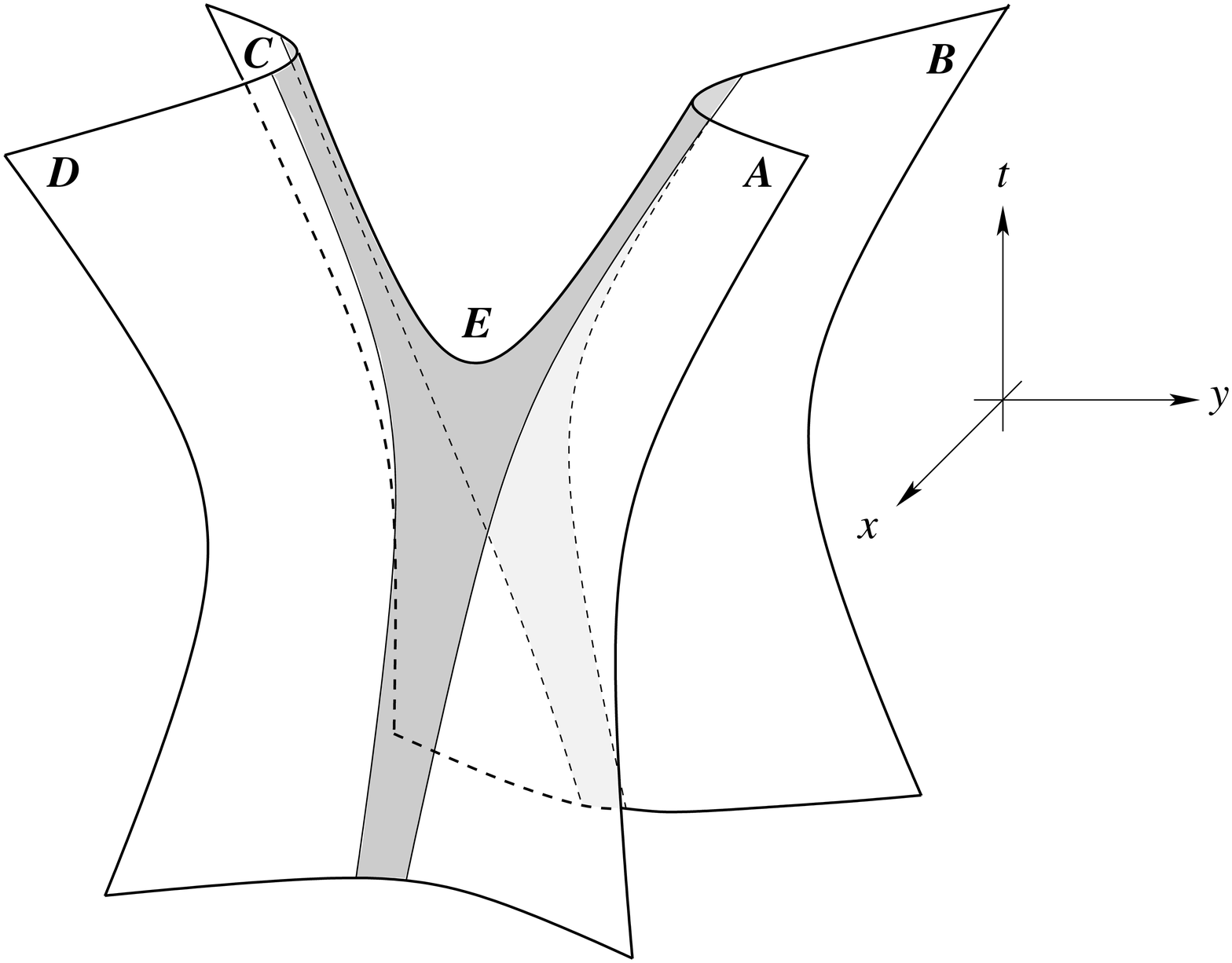}}
\vspace{3mm}
\begin{center}
\begin{minipage}{14cm}
     \caption{\small 
       Generalised Scherk Braneworld. 
       (The $(z,w)$ coordinates are suppressed.) 
       In the grey region $E$, the brane metric becomes Riemannian. 
       The boundaries between the Riemannian region and the four Lorentzian 
       regions are null-like, and there the intrinsic curvature
       diverges. The embedding is, however, smooth across the singularities. 
       } 
        \protect \label{fig:lorentzian-scherk}  
\end{minipage}
\end{center}
\end{figure}

The induced metric is
\ben
ds^2 =   2 dx dy { \sinh x \sinh y \over \cosh x \cosh y } 
    + {dx^2  \over \cosh^2 x } + {dy ^2 \over \cosh ^2 y} 
    +  dz ^2   + dw^2 . 
\label{metric}
\een 
One checks that this is Lorentzian outside the regions
bounded by the four hyperbola shaped 
curves  $\sinh ^2 x \sinh ^2 y >1$, while inside the four curves
there is a connected region including the $x$ and $y$ coordinate
axes in which the metric is positive definite. 
We label the four Lorentzian regions $A,\,B,\,C$, and $D$, 
according to which quadrant they lie in, 
and call the region inside the four curves $E$, 
as shown in Figure~\ref{fig:lorentzian-scherk}.

\begin{figure}[h] 
 \centerline{\epsfxsize = 5cm \epsfbox{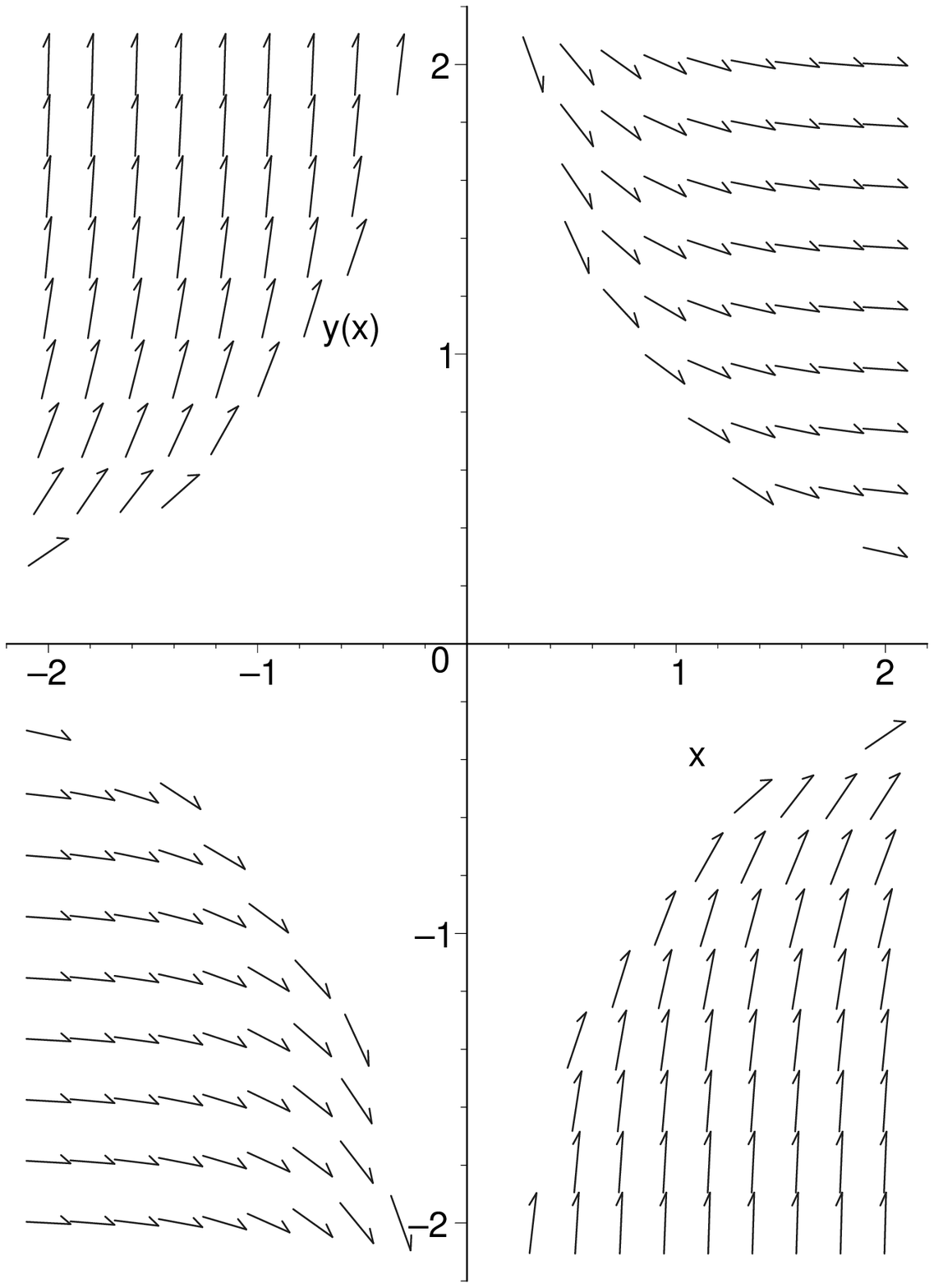}} 
\vspace{3mm}
\begin{center}
\begin{minipage}{14cm}
     \caption{\small Field of null geodesic tangent ``$+$'' projected
       on $(x,y)$-plane.  
       } 
        \protect \label{fig:null-vector-plus}
\end{minipage}
\end{center}
\end{figure}

\begin{figure}[h] 
 \centerline{\epsfxsize = 5cm \epsfbox{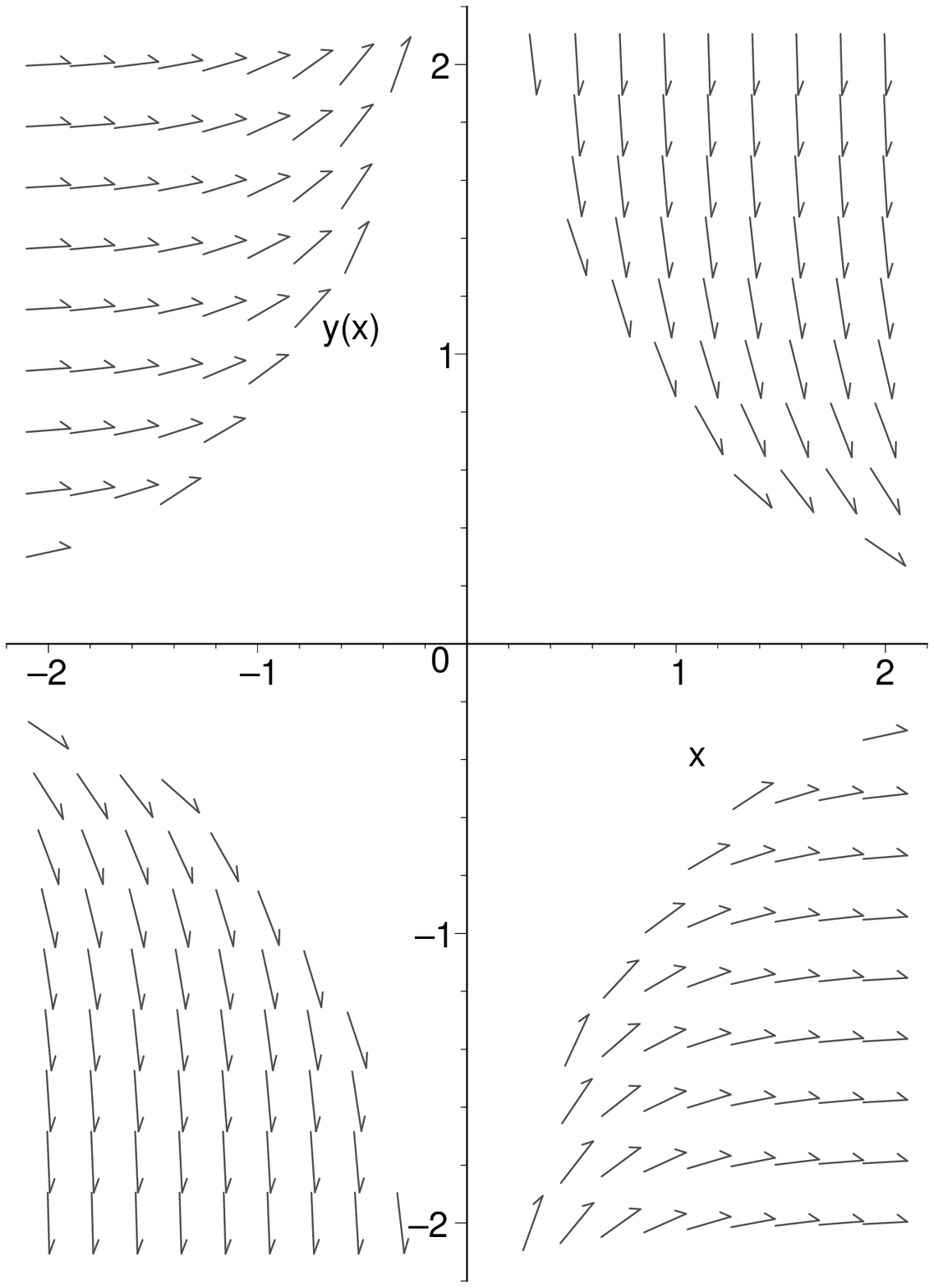}} 
\vspace{3mm}
\begin{center}
\begin{minipage}{14cm}
     \caption{\small Field of null geodesic tangent ``$-$'' projected
       on $(x,y)$-plane. 
       } 
        \protect \label{fig:null-vector-minus}
\end{minipage}
\end{center}
\end{figure}

Let us see the causal structure of this brane. 
The null geodesics with respect to the brane metric obey
\ben
 \left({dy \over dx} \right)_\pm = \frac{\cosh y}{\cosh x}
       \left\{- \sinh x \sinh y \pm \sqrt{\sinh^2 x \sinh^2 y - 1} \right\}.
\een
One can observe that the light cone becomes thinner and thinner as 
the boundary surface is approached 
(See Figure~\ref{fig:null-vector-plus}, ~\ref{fig:null-vector-minus}). 
Since the boundary surface is specified as $\sinh x \sinh y = \pm1$, 
it follows that 
$(dx/\cosh x \pm dy/\cosh y)_{boundary} =0$. 
It then turns out that the induced metric on the boundary 
surface is degenerate. Indeed, it can be checked that 
\ben
 \left({dy \over dx} \right)_{\pm} 
 \left({dy \over dx} \right)^{-1}_{boundary} =1,  
\een
and thus the boundary surface is tangent to the light cone. 
Any causal curves with respect to the brane metric therefore 
cannot hit the corresponding boundary surface, $\partial A$, 
$\partial B$, $\partial C$, $\partial D$. 
Thus, from the view point of the brane causal structure, 
one might think of the boundary as the ``spacelike infinity $i^0$''.  
However, the big difference from the conventional notion of 
the spacelike infinity is that, it is at a ``finite'' location 
in spacelike geodesic distance, and that it is a curvature singularity; 
the scalar curvature with respect to the brane metric,  
\ben
 R = -2 \frac{\cosh^2 x \cosh^2 y}{(1 - \sinh^2 x \sinh^2 y)^2}, 
\een 
diverges there. 

\begin{figure}[h] 
 \centerline{\epsfxsize = 5cm \epsfbox{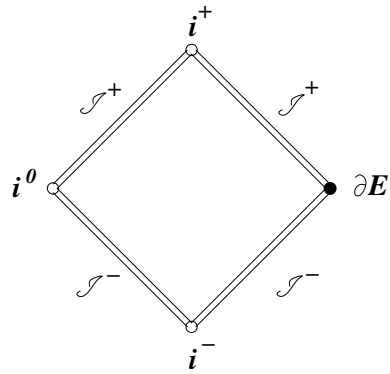}}
\vspace{3mm}
\begin{center}
\begin{minipage}{14cm}
     \caption{\small The causal structure of each Lorentzian region 
        with respect to the brane metric. 
        The $(z,w)$ coordinates are suppressed. 
        $\I^\pm$, $i^\pm$, and $i^0$ denote the infinities of the
        conventional meaning. For example, in the region $A$, 
        $x =+\infty$ at $\I^+$ and $y =+\infty$ at $\I^-$, and 
        in the region $B$, $x =-\infty$ at $\I^+$ and $y =+\infty$ at $\I^-$.  
        $\partial E$ is the curvature
        singularity where spacelike geodesics can reach within finite 
        distance. From the bulk view point, the four copies, $A$, $B$,
        $C$, and $D$ of this panel are smoothly connected via
        $\partial E$. 
       } 
        \protect \label{fig:brane-causality}  
\end{minipage}
\end{center}
\end{figure}

In the asymptotic region $x,y \rightarrow \pm \infty$, 
the coordinates $x$ and $y$ look like null coordinate, and 
the brane metric becomes flat. 
Thus, we have four disconnected Rindler-wedge-like 
$4$-dimensional universes as in Figure~\ref{fig:brane-causality}. 

{}From the bulk view point, the worldsheet of the brane is connected 
and everywhere smooth even at $\partial E$. The four regions $A$, $B$,
$C$, $D$ are connected via $\partial E$.

The region $E$ of the positive definite metric has a finite total area 
\ben
A= \int_E {dx \over \cosh x } { dy \over \cosh y} \sqrt {1 -\sinh^2 x
  \sinh^2 y} < \int^\infty_{-\infty} { dx  \over \cosh x} 
  \int^{ \infty}_{-\infty} { dy \over \cosh y} = { \pi ^2 \over 4}. 
\een 
One may also find that the entire 
$x$ and $y$  axes, which lie entirely within the spacelike region $E$
have finite total length. One may  
 map $x, y \in {\Bbb R} ^2  $ into a square of side $\pi$ by
 introducing coordinates $X,Y \in ( -\pi/2 , \pi / 2) 
 \times (- \pi /2, \pi /2)$ defined by    
\ben
\sinh x= { 1\over \tan X},  \qquad  \sinh y = {1 \over \tan Y}. 
\een 
In these coordinates the metric is
\ben
ds ^2 = d X^2 + d Y ^2 + 2{ d X d Y \over \tan X \tan Y } + dz ^2 + d w^2.  
\een
The $X$ and $Y$ axes correspond to infinity and the opposites sides of
the square, which correspond to the $x$ and $y$ axes, 
must be identified. The spacelike region is now the 
exterior of the curve 
\ben
 \tan^2 X \tan^2  Y=1.
\een

Intrinsically, $E$  is a compact region and the contribution 
it makes to the action (\ref{action}) is of course pure-imaginary 
but nevertheless a finite multiple of the integral over the remaining 
spatial  coordinates $z$ and $w$ 
\ben
  \pm i A T_3 \int dz dw.  
\een      

If we were thinking of a string rather than a 3-brane, the
contribution of $E$ 
to the imaginary part of the action would be finite.

Existence theorems for analogues of Scherk's singly periodic 
minimal $2$-surface in ${\Bbb E}^{3}$ have been given for 
$(n-1)$-ly periodic minimal $n$-surfaces in ${\Bbb E}^{n+1}$ in 
Ref.~\cite{Pacard}. 
Unfortunately no explicit solutions appears to be known and 
we are therefore unable to use it to perform a Wick rotation.  

\subsection{Bulk Field Theory Viewpoint} 

In fact one can model the process of cosmic string commutation
using not the Nambu-Goto action but a classical field theory 
of Nielsen-Olesen type, in other words using the abelian Higgs model
with abelian gauge field $A_\mu$ and complex scalar field $\phi$.
There is a static vortex solution, in which one identifies
the location of the vortex or string  with the zero of the Higgs field
$\phi(x)=0$. In the time dependent case it has been shown 
numerically the commutation takes place~\cite{Shellard}. 
During this process one may attempt to identify the position
of the moving string with the zero of the Higgs. However
the resulting world sheet does not and cannot remain timelike.
Judged by the zero, one would have to say that portions of the string
travel faster than light. However the energy momentum tensor of the
abelian Higgs model satisfies the dominant energy condition.
The energy momentum vector  and hence the flow of energy is
timelike in all frames. It follows from this observation that
the zero of the Higgs does not in general  track 
the distribution of energy any more faithfully than does
the intersection of two almost  parallel searchlight beams
track the distribution of photon energy.

What one is seeing therefore is a breakdown of the distinction
between bulk and brane. In our minimal surface model, the best 
that we can is  to  mimic the very complicated dynamics
of the bulk fields by introducing a Riemannian intermediate
region. However it should not be endowed with any deep 
significance. It is merely a sop to compensate for our loss.
As Minkowski might well have said, four dimensional 
spacetime shall be no more, from now on all that remains
is the higher dimensional bulk spacetime.

\subsection{More Examples of Scherk Braneworlds} 

Signature changing braneworlds discussed above can be embedded 
into some curved target space. Simple examples are given 
by considering a conformally flat bulk with the metric 
\ben 
  g_{\mu \nu}dX^\mu dX^\nu 
  = e^{2\lambda} (-dt^2 + du^adu^a), 
\label{metric:curved}
\een 
where $\lambda$ is a function of $X^\mu$. 
Then, taking 
the Monge-gauge, $X^0=t(u^a)$, $X^a =u^a$ with $a=1,\cdots n=p+1$, 
one can express the equation of motion~(\ref{E-L:eom}) as 
\ben
  \left\{ 1- (\partial_a t) \partial_a t \right\} 
  \left[ \partial^2_b t + n (\partial_b t) \partial_b \lambda
         + n \left\{ 1- (\partial_a t) \partial_a t \right\} 
             \partial_t \lambda 
  \right]  
  + (\partial_a t)(\partial_b t) \partial_a \partial_b t 
  =0 .  
\een
One can find (\ref{soln}) again a solution to the above 
equation of motion, if $\lambda$ is independent of $X^0,\,X^1,\,X^2$. 
The induced metric is given by $e^{2\lambda} $ times 
the metric (\ref{metric}). 
In particular, when $\lambda = -\log z$, it describes a Scherk 
braneworld in anti-de Sitter space.

If one takes a spacelike height function, then one can obtain another 
variant of the solution (\ref{soln}). Consider bulk metric 
(\ref{metric:curved}) with $-dt^2$ replaced by $d\chi^2$ 
and $du^adu^a$ by the Minkowski metric $\eta_{ab}du^adu^b$.  
With the gauge choice $X^a =u^a$, $a=0,1,\cdots,p$ and 
$X^{p+1}=\chi(u^a)$, one has the equation of motion 
\bea
 && 
  \left[ 1 + \eta^{cd} (\partial_c \chi) \partial_d \chi \right] 
  \left[ \eta^{ab}\{ \partial_a\partial_b \chi 
                    - n (\partial_a \chi) \partial_b \lambda \} 
         + n \left\{ 1+ \eta^{cd}(\partial_c \chi) \partial_d \chi \right\} 
         \partial_\chi \lambda 
  \right]  
\non \\
 &&
  - \eta^{ab}\eta^{cd}(\partial_a \chi)(\partial_d \chi) 
                       \partial_b \partial_c \chi  
  =0 .  
\eea  
If $\lambda$ does not depend on $t,\,y,\,\chi$, one can find a
solution 
\ben 
  \chi = \log(\sinh t) - \log(\sin y) , 
\een 
which is analogous to (\ref{soln}). 
The corresponding brane metric (for $p=3$) is 
\ben
   ds^2 = e^{2\lambda} 
            \left(
                 \frac{dt^2}{\sinh^2t} + \frac{dy^2}{\sin^2 y} 
                 - 2 \frac{\cosh t \cos y}{\sinh t \sin y} dtdy
                 + dz^2 + dw^2 
            \right) ,  
\een 
where we have set $u^0=t$, $u^1=y$, $u^2=z$, $u^3=w$. 
Inspecting the determinant of this metric, one see that the 
signature change takes place at $\cosh^2t \cos^2 y=1$. 

\begin{figure}[h] 
 \centerline{\epsfxsize = 8cm \epsfbox{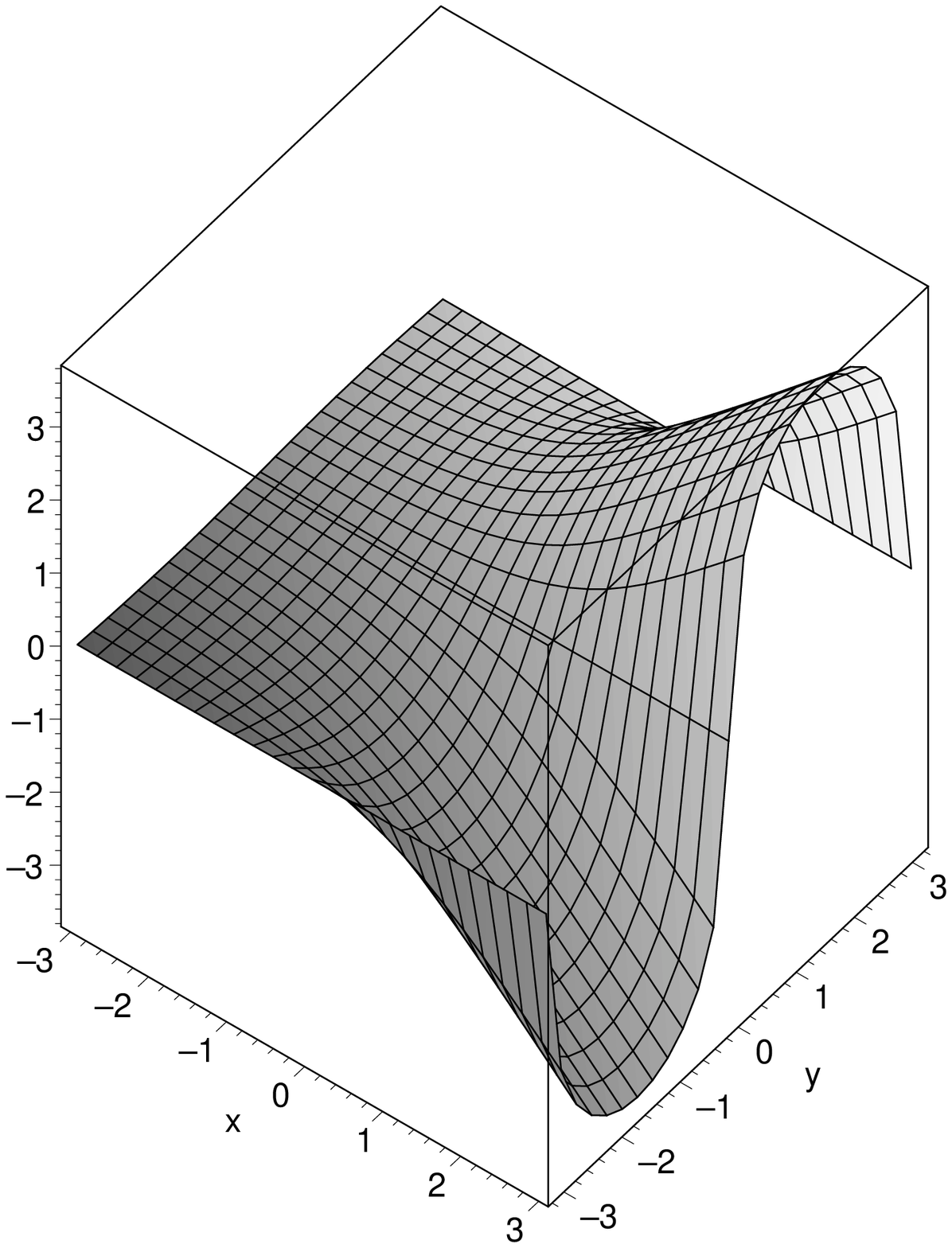}} 
\vspace{3mm}
\begin{center}
\begin{minipage}{14cm}
     \caption{\small 
       A variant of Scherk braneworld, obtained by using 
       a spacelike height function. (The $(z,w)$ coordinates are 
       suppressed.) 
       This solution also describes a signature changing braneworld 
       (Riemannian region in $x<0$ and Lorentzian regions in $x>0$). 
       } 
        \protect \label{fig:variant-Scherk}
\end{minipage}
\end{center}
\end{figure}

The world volume geometry is shown
in Figure~\ref{fig:variant-Scherk}, which may be seen as 
a Lorentzian braneworld (at positive $x$) originated from Euclidean 
region (at negative $x$). 

\begin{figure}[h] 
 \centerline{\epsfxsize = 5cm \epsfbox{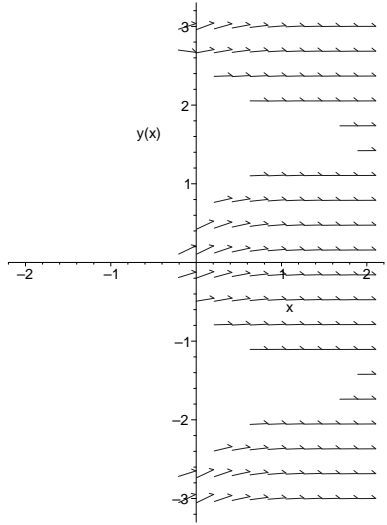}} 
\vspace{3mm}
\begin{center}
\begin{minipage}{14cm}
     \caption{\small Field of null geodesic tangent ``$+$'' projected
       on $(x,y)$-plane. Most regions in $x<0$ are Euclidean. 
       } 
        \protect \label{fig:null-vector-variant-plus}
\end{minipage}
\end{center}
\end{figure}

\begin{figure}[h] 
 \centerline{\epsfxsize = 5cm \epsfbox{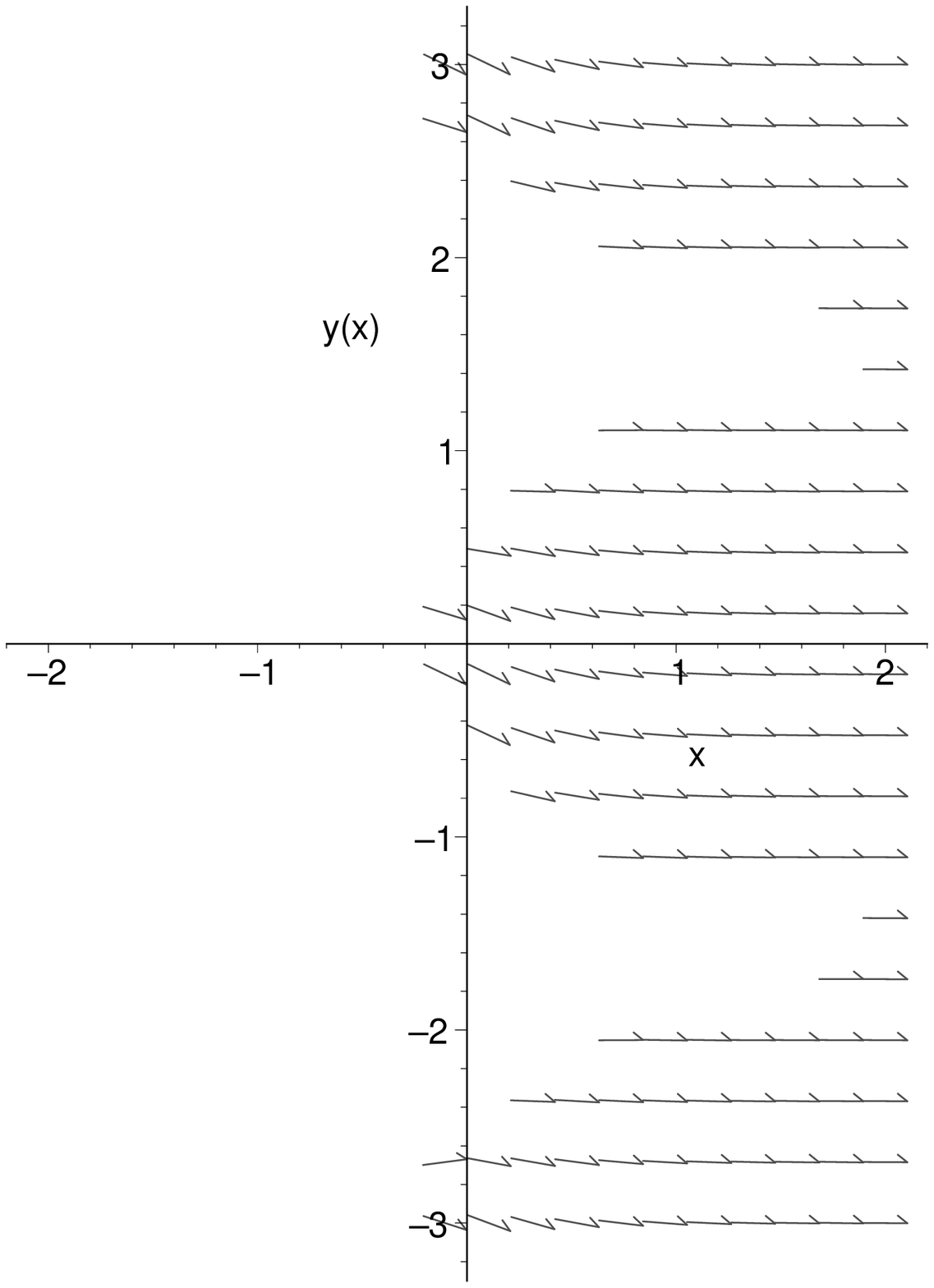}} 
\vspace{3mm}
\begin{center}
\begin{minipage}{14cm}
     \caption{\small Field of null geodesic tangent ``$-$'' projected
       on $(x,y)$-plane. 
       } 
        \protect \label{fig:null-vector-variant-minus}
\end{minipage}
\end{center}
\end{figure}

In terms of $(x,y)$, the null geodesic equation is written by 
\ben
   \left({dy \over dx} \right)_\pm = 
  {e^x\sin y \cos y \pm \sqrt{e^{2x}\sin^2 y \cos^2 y - 1 +e^{2x} 
  -2e^{2x}\sin^2 y } \over 1 - e^{2x} + 2e^{2x}\sin^2 y } . 
\een 
The light cones can be observed from 
Figures~\ref{fig:null-vector-variant-plus}, 
and~\ref{fig:null-vector-variant-minus}, showing that  
the light cones become tangent to the boundary between 
Lorentzian and Euclidean regions.  

One can also find a similar solution, 
$\chi = \log(\sinh t) - \log(\sin y)$. However, 
this solution does not describe signature change. 


\subsection{Generalised Helicoids}   

Another possible sources of interesting examples might come from
Lorentzian version of the generalised helicoids considered 
by~\cite{BarbosadoCarmo}. These are $p+1$ dimensional submanifolds 
$\Sigma _{p+1}$ invariant under a $p$-dimensional translation group 
${\Bbb R}^p$. Such submanifolds are said to be \lq ruled' and the
orbits of the translation group ${\Bbb R}^p$ are flat $p$-planes in 
the ambient flat spacetime. The submanifold $\Sigma _{p+1}$ may be
thought of as a one parameter family of $p$-planes. 
A particularly intriguing case may be obtained from the work 
of section (1.6) of~\cite{BarbosadoCarmo}. 
Let $x(\tau)$ be a timelike curve $\gamma$ 
in ${\Bbb E}^{5,1}$, where $\tau$ is propertime along the curve $\gamma$. 
Let ${\bf e}_a$, $a=0,1,\dots,5$ be a pseudo-orthonormal Frenet frame 
along the curve $\gamma$, so that ${\bf e}_0$ is the unit timelike 
tangent vector to the curve and ${\bf e}_i$, $i=1,\dots,5$ are
spacelike and 
\ben
{d {\bf e}_a \over d \tau} = \omega_a \thinspace {}^b (\tau) 
   {\bf e}_b, 
\label{spin}                
\een
where $\omega_a \thinspace ^c (\tau) \eta _{cb}$ is a skew-symmetric
matrix.      

Now consider the immersion:
\ben
{\bf X} ({u^0} = \tau, u^1, u^2, u^3)  
 = {\bf x}(\tau) + u^1 {\bf e}(\tau)_1 + u^2 {\bf e}(\tau)_3 
                 + u^3 {\bf e}(\tau)_5, 
\label{Gener-Helicoids}
\een
and $d{{\bf x}(\tau)}/d\tau ={\bf e}_0$.  
A simple case is given by 
\ben
 \omega_{01}=-k_0 ,\quad 
 \omega_{12}=k_1 ,\quad 
 \omega_{23}=k_2 ,\quad 
 \omega_{34}=k_3 ,\quad 
 \omega_{45}=k_4 ,\quad 
 \omega_{50}=k_5 .    
\een 
The immersion~(\ref{Gener-Helicoids}) then satisfies 
the equation of motion~(\ref{E-L:eom}) for arbitrary constants 
$k_a$. The induced metric, $h_{ab}$, takes the static form, 
\bea   
 ds^2 &=& -\left[
                 (1-k_0 x-k_5z)^2 -(k_1x-k_2y)^2 -(k_3y-k_4z)^2
           \right] d\tau^2 
\non \\   
 && {} 
  + dx^2 + dy^2 + dz^2 ,  
\eea  
where we have set $u^1=x,\,u^2=y,\,u^3=z$. 
Note that for $k_0 \neq 0$, $k_i=0$, the above metric describes just 
a flat plane in the Rindler coordinates (with the replacement 
$1- k_0 x \rightarrow k_0 x$). 

When $k_0 = \beta \neq 0$, $k_2=\omega$, $k_{1,3,4,5}=0$, 
with appropriate choice of ${\bf e}_a$, one can express the 
immersion as  
\bea
 X &=& \left(x-\frac{1}{\beta}\right)
     \left[-\sinh\beta \tau\frac{\partial}{\partial X^0}  
           + \cosh\beta \tau\frac{\partial}{\partial X^1} 
     \right]
\non \\ 
   &&{}  
    + y \left[ - \sin \omega \tau \frac{\partial}{\partial X^2}  
                 + \cos \omega \tau \frac{\partial}{\partial X^3} 
          \right] 
    + z \frac{\partial}{\partial X^5} . 
\label{expr:immersion}
\eea  
Since $h_{\tau \tau}= \omega^2 y^2-(1-\beta x)^2$, the induced metric 
changes its signature at $\omega y = \pm (1-\beta x)$. 
The Jacobi-matrix is of rank 4 at the points of signature 
change, except the point $x= 1/\beta$ and $y =0$, 
hence the immersion itself is regular there. 
However, the intrinsic curvature, 
\bea
 R=\frac{2}{(h_{\tau\tau})^2}
   [ \beta^4 + \omega^4 +(\beta^2+\omega^2)h_{\tau\tau} ] , 
\eea
of the induced metric diverges there. The signature change therefore 
looks again an occurrence of a curvature singularity from the view 
point of residents on this helicoid, as in the Scherk braneworld case.  

\section{Discussion}  

We have provided models of the braneworlds that admit 
topology change and signature change in a smooth Lorentzian 
bulk. We also gave models of oscillating and spinning brane universes 
by generalising Lorentzian catenoid and helicoid. 
Our braneworld models obey the Dirac-Born-Infeld equations of motion, 
but their self-gravity was neglected so as to allow a simple model.  


Concerning the signature changing brane models, 
we should point out that although 
there are certain similarities with the minimal
surfaces we have used in this paper and
certain instantons, i.e. complex saddle points used to
describe the high energy limit of string scattering 
\cite{Gross1, Gross2, GrossMende1, GrossMende2}
which were extended to include the presence
of D-branes in~\cite{BachasPioline} 
there are important differences. Firstly we have in mind 
not only strings, i.e., $p=1$ but the case of general $p$-branes, 
$p>1$. Indeed for us the most interesting case is $p=3$. 
Secondly, our solutions are real, where as those used 
in~\cite{Gross1, Gross2, GrossMende1, GrossMende2} 
are pure imaginary although the induced metric is
always real but positive definite. It is possible in some cases, 
for example, that some forms of Scherks' surface may be related 
by analytic continuation to those used 
in~\cite{Gross1, Gross2, GrossMende1, GrossMende2,BachasPioline}. 
However, the scattering processes we have had in mind are, 
at least from the bulk point of view, entirely classical.
As we have speculated above, it may be that from the world volume 
point of view they may be thought of in a more quantum mechanical
way. If so, the differences in our approach and that of 
\cite{Gross1,Gross2,GrossMende1,GrossMende2,BachasPioline} 
may not be so great as at first appears. It would clearly be 
of great interest to pursue this connection further.  
A treatment of intersecting $D$-branes in relation to 
tachyon condensation is given in Ref.~\cite{HashimotoNgaoka}.

In our examples, the boundary between the Lorentzian and Euclidean
regions corresponds to a curvature singularity with respect 
to the induced metric of a signature changing braneworld. 
In order for inhabitants on the brane to understand the situation, 
they need to develop quantum gravity theory in 4-dimensions. 
On the other hand, from the bulk view point, 
the brane is everywhere, even at the points corresponding to the 
singularity, smooth. 
It can be simply described by embedding equation. 
This observation suggests an possible way of resolving spacetime 
singularities in the braneworld context. 
This also conforms to the spirit of holographic principle or 
bulk-boundary correspondence, in the sense that quantum theory on 
a brane could be understood in terms of bulk classical theory. 
In order to make spacetime metric real, the Euclidean approach 
needs the existence of a totally geodesic spacelike hypersurface. 
This is one of great limitations of the uses of the Euclidean approach. 
In the present model, the Euclidean region of a braneworld is connected 
with Lorentzian region at spacelike surfaces which 
are not totally geodesic surface but correspond to a singularity. 
It would be interesting if one can develop Euclidean Quantum Gravity 
by implementing the signature change of a braneworld in a smooth 
Lorentzian bulk spacetime.

\bigskip 
\begin{center}
{\bf Acknowledgements}
\end{center}
GWG would like to thank Costas Bachas, Michael Green, Ergin Sezgin and 
Andrew Strominger for helpful discussions and encouragement. 
In particular, conversations with Costas Bachas at ENS in Paris 
during late 1999 were extremely illuminating. A preliminary version 
of this work was described at the George and Cynthia Mitchell 
Institute for Fundamental Physics in Spring 2002. 
We also wish to thank Koji Hashimoto, Jose Senovilla, 
Paul Shellard and Supriya Kar. 
This research was supported in part by the Japan Society for 
the Promotion of Science.

\end{document}